\RequirePackage{lineno}
\documentclass[aps,prl,twocolumn,superscriptaddress,showpacs,preprintnumbers,amsmath,amssymb]{revtex4}

\usepackage{color}
\usepackage{longtable}
\usepackage{graphicx} % Include figure files
\usepackage{dcolumn}  % Align table columns on decimal point
\usepackage{pdflscape}
\usepackage{lineno}

\begin{document}

\vspace*{-3\baselineskip}
%\resizebox{!}{3cm}{\includegraphics{belle.eps}}

\preprint{\vbox{
    %\hbox{Version 0.3}
    \hbox{Belle Preprint 2017-03}
    \hbox{KEK Preprint 2016-64}
%                \hbox{Belle-CONF-1609}
}}
 
\title{\quad\\[0.5cm] Search for {\boldmath$CP$} Violation and Measurement of the Branching Fraction in the Decay {\boldmath$D^{0} \to K^0_S K^0_S$}}
%\title{\quad\\[0.5cm]  Measurement of branching fraction and {\boldmath$CP$} violation in the decay {\boldmath$D^{0} \to K^0_S K^0_S$}}

\noaffiliation
%%%\affiliation{Aligarh Muslim University, Aligarh 202002}
\affiliation{University of the Basque Country UPV/EHU, 48080 Bilbao}
\affiliation{Beihang University, Beijing 100191}
\affiliation{University of Bonn, 53115 Bonn}
\affiliation{Budker Institute of Nuclear Physics SB RAS, Novosibirsk 630090}
\affiliation{Faculty of Mathematics and Physics, Charles University, 121 16 Prague}
%%%\affiliation{Chiba University, Chiba 263-8522}
\affiliation{Chonnam National University, Kwangju 660-701}
\affiliation{University of Cincinnati, Cincinnati, Ohio 45221}
\affiliation{Deutsches Elektronen--Synchrotron, 22607 Hamburg}
\affiliation{University of Florida, Gainesville, Florida 32611}
\affiliation{Department of Physics, Fu Jen Catholic University, Taipei 24205}
\affiliation{Justus-Liebig-Universit\"at Gie\ss{}en, 35392 Gie\ss{}en}
%%%\affiliation{Gifu University, Gifu 501-1193}
%%%\affiliation{II. Physikalisches Institut, Georg-August-Universit\"at G\"ottingen, 37073 G\"ottingen}
\affiliation{SOKENDAI (The Graduate University for Advanced Studies), Hayama 240-0193}
%%%\affiliation{Gyeongsang National University, Chinju 660-701}
\affiliation{Hanyang University, Seoul 133-791}
\affiliation{University of Hawaii, Honolulu, Hawaii 96822}
\affiliation{High Energy Accelerator Research Organization (KEK), Tsukuba 305-0801}
\affiliation{J-PARC Branch, KEK Theory Center, High Energy Accelerator Research Organization (KEK), Tsukuba 305-0801}
%%%\affiliation{Hiroshima Institute of Technology, Hiroshima 731-5193}
\affiliation{IKERBASQUE, Basque Foundation for Science, 48013 Bilbao}
%%%\affiliation{University of Illinois at Urbana-Champaign, Urbana, Illinois 61801}
\affiliation{Indian Institute of Science Education and Research Mohali, SAS Nagar, 140306}
\affiliation{Indian Institute of Technology Bhubaneswar, Satya Nagar 751007}
\affiliation{Indian Institute of Technology Guwahati, Assam 781039}
\affiliation{Indian Institute of Technology Madras, Chennai 600036}
\affiliation{Indiana University, Bloomington, Indiana 47408}
\affiliation{Institute of High Energy Physics, Chinese Academy of Sciences, Beijing 100049}
\affiliation{Institute of High Energy Physics, Vienna 1050}
\affiliation{Institute for High Energy Physics, Protvino 142281}
%%%\affiliation{Institute of Mathematical Sciences, Chennai 600113}
\affiliation{INFN - Sezione di Napoli, 80126 Napoli}
\affiliation{INFN - Sezione di Torino, 10125 Torino}
\affiliation{Advanced Science Research Center, Japan Atomic Energy Agency, Naka 319-1195}
\affiliation{J. Stefan Institute, 1000 Ljubljana}
\affiliation{Kanagawa University, Yokohama 221-8686}
\affiliation{Institut f\"ur Experimentelle Kernphysik, Karlsruher Institut f\"ur Technologie, 76131 Karlsruhe}
%%%\affiliation{Kavli Institute for the Physics and Mathematics of the Universe (WPI), University of Tokyo, Kashiwa 277-8583}
\affiliation{Kennesaw State University, Kennesaw, Georgia 30144}
\affiliation{King Abdulaziz City for Science and Technology, Riyadh 11442}
\affiliation{Department of Physics, Faculty of Science, King Abdulaziz University, Jeddah 21589}
\affiliation{Korea Institute of Science and Technology Information, Daejeon 305-806}
\affiliation{Korea University, Seoul 136-713}
\affiliation{Kyoto University, Kyoto 606-8502}
\affiliation{Kyungpook National University, Daegu 702-701}
\affiliation{\'Ecole Polytechnique F\'ed\'erale de Lausanne (EPFL), Lausanne 1015}
\affiliation{P.N. Lebedev Physical Institute of the Russian Academy of Sciences, Moscow 119991}
\affiliation{Faculty of Mathematics and Physics, University of Ljubljana, 1000 Ljubljana}
\affiliation{Ludwig Maximilians University, 80539 Munich}
%%%\affiliation{Luther College, Decorah, Iowa 52101}
%%%\affiliation{University of Malaya, 50603 Kuala Lumpur}
\affiliation{University of Maribor, 2000 Maribor}
\affiliation{Max-Planck-Institut f\"ur Physik, 80805 M\"unchen}
\affiliation{School of Physics, University of Melbourne, Victoria 3010}
%%%\affiliation{Middle East Technical University, 06531 Ankara}
%%%\affiliation{University of Miyazaki, Miyazaki 889-2192}
\affiliation{Moscow Physical Engineering Institute, Moscow 115409}
\affiliation{Moscow Institute of Physics and Technology, Moscow Region 141700}
\affiliation{Graduate School of Science, Nagoya University, Nagoya 464-8602}
\affiliation{Kobayashi-Maskawa Institute, Nagoya University, Nagoya 464-8602}
%%%\affiliation{Nara University of Education, Nara 630-8528}
\affiliation{Nara Women's University, Nara 630-8506}
\affiliation{National Central University, Chung-li 32054}
\affiliation{National United University, Miao Li 36003}
\affiliation{Department of Physics, National Taiwan University, Taipei 10617}
\affiliation{H. Niewodniczanski Institute of Nuclear Physics, Krakow 31-342}
\affiliation{Nippon Dental University, Niigata 951-8580}
\affiliation{Niigata University, Niigata 950-2181}
%%%\affiliation{University of Nova Gorica, 5000 Nova Gorica}
\affiliation{Novosibirsk State University, Novosibirsk 630090}
\affiliation{Osaka City University, Osaka 558-8585}
%%%\affiliation{Osaka University, Osaka 565-0871}
\affiliation{Pacific Northwest National Laboratory, Richland, Washington 99352}
%%%\affiliation{Panjab University, Chandigarh 160014}
%%%\affiliation{Peking University, Beijing 100871}
\affiliation{University of Pittsburgh, Pittsburgh, Pennsylvania 15260}
\affiliation{Punjab Agricultural University, Ludhiana 141004}
%%%\affiliation{Research Center for Electron Photon Science, Tohoku University, Sendai 980-8578}
%%%\affiliation{Research Center for Nuclear Physics, Osaka University, Osaka 567-0047}
\affiliation{Theoretical Research Division, Nishina Center, RIKEN, Saitama 351-0198}
%%%\affiliation{RIKEN BNL Research Center, Upton, New York 11973}
%%%\affiliation{Saga University, Saga 840-8502}
\affiliation{University of Science and Technology of China, Hefei 230026}
%%%\affiliation{Seoul National University, Seoul 151-742}
%%%\affiliation{Shinshu University, Nagano 390-8621}
\affiliation{Showa Pharmaceutical University, Tokyo 194-8543}
\affiliation{Soongsil University, Seoul 156-743}
%%%\affiliation{University of South Carolina, Columbia, South Carolina 29208}
\affiliation{Stefan Meyer Institute for Subatomic Physics, Vienna 1090}
\affiliation{Sungkyunkwan University, Suwon 440-746}
\affiliation{School of Physics, University of Sydney, New South Wales 2006}
\affiliation{Department of Physics, Faculty of Science, University of Tabuk, Tabuk 71451}
\affiliation{Tata Institute of Fundamental Research, Mumbai 400005}
\affiliation{Excellence Cluster Universe, Technische Universit\"at M\"unchen, 85748 Garching}
\affiliation{Department of Physics, Technische Universit\"at M\"unchen, 85748 Garching}
\affiliation{Toho University, Funabashi 274-8510}
%%%\affiliation{Tohoku Gakuin University, Tagajo 985-8537}
\affiliation{Department of Physics, Tohoku University, Sendai 980-8578}
\affiliation{Earthquake Research Institute, University of Tokyo, Tokyo 113-0032}
\affiliation{Department of Physics, University of Tokyo, Tokyo 113-0033}
\affiliation{Tokyo Institute of Technology, Tokyo 152-8550}
\affiliation{Tokyo Metropolitan University, Tokyo 192-0397}
%%%\affiliation{Tokyo University of Agriculture and Technology, Tokyo 184-8588}
\affiliation{University of Torino, 10124 Torino}
%%%\affiliation{Toyama National College of Maritime Technology, Toyama 933-0293}
%%%\affiliation{Utkal University, Bhubaneswar 751004}
\affiliation{Virginia Polytechnic Institute and State University, Blacksburg, Virginia 24061}
\affiliation{Wayne State University, Detroit, Michigan 48202}
\affiliation{Yamagata University, Yamagata 990-8560}
\affiliation{Yonsei University, Seoul 120-749}
% \author{A.~Abdesselam}\affiliation{Department of Physics, Faculty of Science, University of Tabuk, Tabuk 71451} % Tabuk
  \author{N.~Dash}\affiliation{Indian Institute of Technology Bhubaneswar, Satya Nagar 751007} % IITB
  \author{S.~Bahinipati}\affiliation{Indian Institute of Technology Bhubaneswar, Satya Nagar 751007} % IITB
  \author{V.~Bhardwaj}\affiliation{Indian Institute of Science Education and Research Mohali, SAS Nagar, 140306} % IISERM
  \author{K.~Trabelsi}\affiliation{High Energy Accelerator Research Organization (KEK), Tsukuba 305-0801}\affiliation{SOKENDAI (The Graduate University for Advanced Studies), Hayama 240-0193} % KEK
  \author{I.~Adachi}\affiliation{High Energy Accelerator Research Organization (KEK), Tsukuba 305-0801}\affiliation{SOKENDAI (The Graduate University for Advanced Studies), Hayama 240-0193} % KEK
% \author{K.~Adamczyk}\affiliation{H. Niewodniczanski Institute of Nuclear Physics, Krakow 31-342} % Krakow
  \author{H.~Aihara}\affiliation{Department of Physics, University of Tokyo, Tokyo 113-0033} % Tokyo
  \author{S.~Al~Said}\affiliation{Department of Physics, Faculty of Science, University of Tabuk, Tabuk 71451}\affiliation{Department of Physics, Faculty of Science, King Abdulaziz University, Jeddah 21589} % Tabuk
% \author{K.~Arinstein}\affiliation{Budker Institute of Nuclear Physics SB RAS, Novosibirsk 630090}\affiliation{Novosibirsk State University, Novosibirsk 630090} % BINP
% \author{Y.~Arita}\affiliation{Graduate School of Science, Nagoya University, Nagoya 464-8602} % Nagoya
  \author{D.~M.~Asner}\affiliation{Pacific Northwest National Laboratory, Richland, Washington 99352} % PNNL
% \author{T.~Aso}\affiliation{Toyama National College of Maritime Technology, Toyama 933-0293} % Toyama
% \author{H.~Atmacan}\affiliation{University of South Carolina, Columbia, South Carolina 29208} % SouthCarolina
  \author{V.~Aulchenko}\affiliation{Budker Institute of Nuclear Physics SB RAS, Novosibirsk 630090}\affiliation{Novosibirsk State University, Novosibirsk 630090} % BINP
  \author{T.~Aushev}\affiliation{Moscow Institute of Physics and Technology, Moscow Region 141700} % MIPT
% ...
% Ljubljana
% \author{N.~Zwahlen}\affiliation{\'Ecole Polytechnique F\'ed\'erale de Lausanne (EPFL), Lausanne 1015} % Lausanne
\collaboration{The Belle Collaboration}

\begin{abstract}
We report a study of the decay $D^0 \to K^0_S K^0_S$ using 921~fb$^{-1}$ of data collected at or near the $\Upsilon(4S)$ and $\Upsilon(5S)$ resonances with the Belle detector at the KEKB asymmetric energy $e^+e^-$ collider. The measured time-integrated $CP$ asymmetry is $ A_{CP}(D^0 \to K^0_S K^0_S) = (-0.02 \pm 1.53 \pm 0.02 \pm 0.17) \%$, and the branching fraction is $\mathcal{B} (D^{0}\rightarrow K_{S}^{0}K_{S}^{0})$ = (1.321 $\pm$ 0.023  $\pm$ 0.036 $\pm$ 0.044) $\times$ 10$^{-4}$, where the first uncertainty is statistical, the second is systematic, and the third is due to the normalization mode ($D^0 \to K_S^0 \pi^0$). These results are significantly more precise than previous measurements available for this mode. The $A_{CP}$ measurement is consistent with the standard model expectation.

\end{abstract}

\maketitle

%.....................................................................

Charge-parity violation (CPV) in charm meson decays has not yet been observed and is predicted to be small [$\mathcal{O}(10^{-3})$] in the standard model (SM)~\cite{grossman}. Hence, an observation of larger CPV in charm decays could be interpreted as a sign of new physics (NP)~\cite{grossman}. Singly Cabibbo-suppressed (SCS) decays~\cite{hiller} are of special interest as possible interference with NP amplitudes could lead to large nonzero CPV.  The $D^{0} \to K^0_{S}K^0_{S}$ decay is the most promising channel amongst the SCS decays, as the $CP$ asymmetry may be enhanced to an observable level within the SM, thanks to the interference of the transitions $c\bar{u} \to \bar{s}s$ and $c\bar{u} \to \bar{d}d$, both of which involve the tree-level exchange of a $W$ boson~\cite{Nierste:2015zra}. 

 Assuming the total decay width to be the same for particles and antiparticles, the time-integrated $CP$ asymmetry is defined as
 
 \begin{equation}
   A_{CP}=\frac{\Gamma (D^{0}\rightarrow K_{S}^{0}K_{S}^{0})-\Gamma (\bar D^{0}\rightarrow K_{S}^{0}K_{S}^{0})}{\Gamma (D^{0}\rightarrow K_{S}^{0}K_{S}^{0})+\Gamma (\bar D^{0}\rightarrow K_{S}^{0}K_{S}^{0})},
\end{equation}

\noindent where $\Gamma$ represents the partial decay width. This asymmetry has three contributions: \\

\begin{equation}
  A_{CP} = A_{CP}^{d} + A_{CP}^{m} + A_{CP}^{i}, 
\end{equation} 

\noindent where $A_{CP}^{d}$ is due to direct CPV (which is decay-mode dependent), $A_{CP}^{m}$ to CPV in $D^0$--$\bar{D}{}^0$ mixing, and $A_{CP}^{i}$ to CPV in the interference between decays with and without mixing. The last two terms are independent of the decay final states and are related to the lifetime ($\tau$) asymmetry~\cite{alambda}, 

\begin{equation}
  A_{\Gamma}=\frac{\tau(D^{0})-\tau(\bar D^{0})}{\tau(D^{0})+\tau(\bar D^{0})} = -(A_{CP}^{m} + A_{CP}^{i})\,.
\end{equation}

The world average for $A_{\Gamma} $, $(-0.032\pm 0.026)\%$, is consistent with zero~\cite{hfag}. In the SM, indirect CPV ($A_{CP}^{m} + A_{CP}^{i}$) is expected to be very small, of the order of $10^{-3}$~\cite{grossman}. Direct CPV in SCS decays is further parametrically suppressed [$\mathcal{O}(10^{-4})$], since it arises from the interference of the tree and penguin amplitudes~\cite{brod}. However, these decays, unlike Cabibbo favored or doubly Cabibbo suppressed ones, are sensitive to new SM contributions from strong penguin operators, especially from chromomagnetic dipole operators~\cite{grossman}. A recent SM-based calculation obtains a 95\% confidence level upper limit of 1.1\% for direct $CP$ violation in this decay~\cite{Nierste:2015zra}.

     The search for  time-integrated $CP$ asymmetry in $D^{0} \to K^0_{S} K^0_{S}$ was first performed by CLEO~\cite{Bonvicini:2000qm} using a data sample of 13.7 fb$^{-1}$ of $e^+e^-$ collisions at the $\Upsilon(4S)$ resonance with a measured $CP$ asymmetry of $(-23 \pm 19)\%$. LHCb subsequently measured the same quantity as ($-$2.9 $\pm$ 5.2 $\pm$ 2.2)\%~\cite{Aaij:2015fua}. Both results are consistent with no CPV, in agreement with the SM expectation. Recently, BESIII reported a $D^{0}\rightarrow K^0_{S}K^0_{S}$ branching fraction of $(1.67 \pm 0.11 \pm 0.11)\times$ 10$^{-4}$~\cite{BESIII} by analyzing data corresponding to an integrated luminosity of 2.93 fb$^{-1}$ taken at the $\psi(3770)$ resonance. Belle can significantly improve these measurements using the high-statistics data samples at or near the $\Upsilon(4S)$ and $\Upsilon(5S)$ resonances.

%....................................................................
In this Letter, we measure the branching fraction and the time-integrated $CP$ asymmetry ($A_{CP}$) of the neutral charmed meson decay $D^{0}\rightarrow K_{S}^{0}K_{S}^{0}$. The analysis is based on a data sample that corresponds to an integrated luminosity of 921~fb$^{-1}$ collected with the Belle detector~\cite{abashian} at the KEKB asymmetric-energy $e^+e^-$ collider~\cite{kurokawa} operating at or slightly below the $\Upsilon$(4S) resonance and at the $\Upsilon$(5S) resonance with integrated luminosities of 710.5, 89.2, and 121.4~fb$^{-1}$, respectively. The Belle detector is a large-solid-angle spectrometer, which includes a silicon vertex detector (SVD), a 50-layer central drift chamber (CDC), an array of aerogel threshold Cherenkov counters (ACC), time-of-flight (TOF) scintillation counters, and an electromagnetic calorimeter (ECL) comprised of CsI(Tl) crystals located inside a superconducting solenoid coil that provides a 1.5~T magnetic field. An iron flux return located outside the coil is instrumented to detect $K^{0}_{L}$ mesons and identify muons.

%....................................................................

 For this analysis, the $D^0$ meson is required to originate from the decay $D^{*+} \to D^0 \pi_{s}^+$, where $ \pi_{s}^+$ is a slow pion, in order to identify the $D^0$ flavor and suppress the combinatorial background. 

 The measured raw asymmetry is

\begin{equation}
  A_{\rm raw} = \frac{N(D^0) - N(\bar{D}^0)}{N(D^0) + N(\bar{D}^0)} = A_{CP} + A_{\textrm FB} + A_{\epsilon}^{\pm} + A_{\epsilon}^{K}\,,
\end{equation}

\noindent where all terms are small ($< 1 \%$): $A_{\textrm FB}$ is the forward-backward production asymmetry of $D^0$ mesons, $A^{\pm{}}_{\epsilon}$ is the asymmetry due to different detection efficiencies for positively and negatively charged pions, and $A^{K}_{\epsilon}$ is the asymmetry originating from the distinct strong interaction of $K^0$ and ${\bar{K}}^0$ mesons with nucleons in the detector material. $A_{\textrm FB}$ and $A^{\pm{}}_{\epsilon}$ can be eliminated through a relative measurement of $A_{CP}$ with respect to the well-measured mode $D^0 \to K_S^0 \pi^0$.
The value of $A^{K}_{\epsilon}$ is estimated to be $-0.11\%$ due to a nonvanishing asymmetry originating from the different nuclear interaction of $K^0$ and ${\bar{K}}^0$ mesons with the detector material, estimated in Ref.~\cite{belle_K}.
The $CP$ asymmetry of the signal mode is then expressed as 
\begin{eqnarray}
A_{CP}(D^0 \to K_S^0K_S^0) &=& A_{\rm raw}(D^0 \to K_S^0K_S^0) -\nonumber
 \\
& & A_{\rm raw}(D^0 \to K_S^0\pi^0) + \nonumber
\\
& & A_{CP}(D^0 \to K_S^0\pi^0) + A^{K}_{\epsilon},
\end{eqnarray}
where $A_{CP}(D^0 \to K_S^0\pi^0) = (-0.20 \pm 0.17)\%$~\cite{pdg} is the world-average $CP$ asymmetry of the normalization mode.

The $D^{*+}$ mesons originate mostly from the $e^+e^- \rightarrow c\bar{c}$ process via hadronization, where the inclusive yield has a large uncertainty of 12.5\%~\cite{pdg}. To avoid this uncertainty, we measure the $D^0 \rightarrow K_S^0K_S^0$ branching fraction with respect to that of the $D^0 \rightarrow K_S^0\pi^0$ mode using the following relation:
 
 \begin{equation}\label{eq:9}
 \frac{\mathcal{B}(D^0 \rightarrow K_S^0K_S^0)}{\mathcal{B}(D^0 \rightarrow K_S^0\pi^0)} =   \frac{(N/\epsilon)_{D^0 \rightarrow K_S^0K_S^0}}{(N/\epsilon)_{D^0 \rightarrow K_S^0\pi^0}}. 
  \end{equation}
  Here, $\cal{B}$ is the branching fraction, $N$ is the extracted signal yield, and $\epsilon$ is the reconstruction efficiency. The world-average value of ${\mathcal{B}(D^0 \rightarrow K_S^0\pi^0)}$ = $(1.20 \pm 0.04)\%$ is used~\cite{pdg}. In this ratio, the systematic uncertainties common to the signal and normalization channels cancel. 

The analysis procedure is developed using Monte Carlo (MC) simulation based on events generated using \textsc{EvtGen}~\cite{evtgen}, which includes final-state radiation effects via \textsc{PHOTOS}~\cite{photos}; the detector response is simulated by \textsc{GEANT3}~\cite{geant}. The selection criteria are optimized using a figure of merit defined as $N_{\rm sig}/\sqrt{N_{\rm sig} + N_{\rm bkg}}$, where $N_{\rm sig}$ ($N_{\rm bkg}$) is the number of signal (background) events in the signal region defined as $0.144$~GeV/$c^{2} $ $ < \Delta M < 0.147$~GeV/$c^{2}$ and $1.847$~GeV/$c^{2} $ $ < M (D^{0}) < 1.882$~GeV/$c^{2}$, where $\Delta M = M(D^{*}) - M(D^{0})$, and $M$ is the reconstructed invariant mass of the corresponding meson candidate. We use a signal MC sample with about 400 times more events than expected in the data and estimate $N_{\rm sig}$ assuming ${\cal B}(D^0 \to K_S^0 K_S^0) = 1.8 \times 10^{-4}$~\cite{pdg}. The MC sample used to estimate the background comprises $B{\bar{B}}$ and $q{\bar{q}}$ events, where $q=u,d,s,c$ and corresponds to an integrated luminosity of 6 times that of the data. The background contribution is scaled by the ratio of the number of events in data and MC estimations in the $\Delta M$ sideband defined as $0.148$~GeV/$c^{2} $ $ < \Delta M < 0.160$~GeV/$c^{2}$.

We require a slow pion ($\pi_{s}$) candidate to originate from near the interaction point (IP) by restricting its impact parameters along and perpendicular to the $z$ axis to be less than 3 and 1~cm, respectively. The $z$ axis is defined as the direction opposite the $e^+$ beam. We require that the ratio of the particle identification (PID) likelihoods, ${\cal L}_{\pi}/({\cal L}_{\pi} + {\cal L}_{K})$, be greater than 0.4. Here, ${\cal L}_{\pi}$ (${\cal L}_{K}$) is the likelihood of a track being a pion (kaon) and is calculated using specific ionization from the CDC, time-of-flight information from the TOF scintillation counters and the number of photoelectrons in the ACC. With the above PID requirement, the pion identification efficiency is above $95\%$ with a kaon misidentification probability below $5\%$.

The $K_{S}^{0}$ candidates are reconstructed from pairs of oppositely charged tracks, both treated as pions, and are identified with a neural network (NN)~\cite{neurobayes}. The NN uses the following seven variables: the $K^0_S$ momentum in the laboratory frame, the distance along the $z$ axis between the two track helices at their closest approach, the flight length in the $x$-$y$ plane, the angle between the $K_S^0$ momentum and the vector joining the IP to the $K_S^0$ decay vertex, the angle between the pion momentum and the laboratory-frame direction in the $K_S^0$ rest frame, the distances of closest approach in the $x$-$y$ plane between the IP and the two pion helices, and the total number of hits (in the CDC and SVD) for each pion track. We also require that the reconstructed invariant mass be within $\pm 15$~MeV/$c^{2}$ (about 4 times the resolution) of the nominal $K_{S}^{0}$ mass~\cite{pdg}. The $K_S^0$ reconstruction efficiency is 81.9$\%$. We reconstruct neutral pion candidates from pairs of electromagnetic showers in the ECL that are not matched to any charged track. Showers in the barrel (end-cap) region of the ECL must exceed 60 (100) MeV to be considered as a $\pi^0$ daughter candidate~\cite{piz}. The invariant mass of the $\pi^0$ candidate must lie within $\pm 25$~MeV/$c^2$ (about 4 times the resolution) of the known $\pi^0$ mass~\cite{pdg}. The $\pi^0$ momentum is required to be greater than 640 MeV/$c$. 

To reconstruct $D^{0}$ candidates, we combine two reconstructed $K_{S}^{0}$ candidates for the signal mode (one $K^0_S$ and one $\pi^0$ for the normalization mode) and retain those having an invariant mass in the range $1.847 $~GeV/$c^{2} $ $ < M (D^{0}) < 1.882 $~GeV/$c^2$ [$1.758$~GeV/$c^{2} $ $ < M (D^{0}) < 1.930$~GeV/$c^2$], within $\pm 3 \sigma$ of the nominal $D^{0}$ mass~\cite{pdg}. Finally, $\pi_{s}$ candidates are combined with the $D^{0}$ candidates to form $D^{*}$ candidates, with the requirement that $\Delta M$ lies in the range [0.140, 0.160]~GeV/$c^{2}$. The slow pion is constrained to originate from the IP in order to improve the $\Delta M$ resolution. We require $D^{*+}$ candidates to have a momentum greater than 2.2~GeV/$c$ in the center-of-mass frame. This requirement significantly reduces background from random $D^{0} \pi^{+}_{s}$ combinations.

After all selection criteria, the fraction of signal events with multiple $D^{*}$ candidates is 8.6$\%$. If this is due to multiple $D^{0}$ candidates, we retain the one having the smallest $\sum \chi^{2}_{K_{S}^{0}}$ , where $\chi^{2}_{K_{S}^{0}}$ is the test statistic of the $K_{S}^{0}$ vertex-constraint fit. In case several $D^{*}$ candidates remain, the one having the charged pion with the smallest transverse impact parameter is retained. This choice correctly identifies the true $D^{*} \to D^0 [K_S^0 K_S^0]\, \pi_{s}$ decay with an efficiency of $98\%$. The best-candidate selection efficiency is the same for $D^{*+}$ and $D^{*-}$ candidates. For the normalization mode, the fraction of signal events with multiple $D^{*}$ candidates is 27.3$\%$. If this is due to multiple $D^{0}$ candidates, we retain the one having the smallest value for the sum of $\chi^{2}_{K_{S}^{0}}$ and $\chi^{2}_{\pi^{0}}$, where $\chi^{2}_{\pi^{0}}$ is the test statistic of the ${\pi^{0}}$ mass-constraint fit. This procedure for $D^0 \to K_S^0 \pi^0$ selects the correct candidate with an efficiency of $89\%$.

%\begin{figure}
%\begin{center}
%\includegraphics[width=8.5cm]{./kspi0_tot_8.eps} 
%\includegraphics[width=8.5cm]{./ksks_tot_8.eps}
%\caption{Distributions of the mass difference $\Delta M$ for the $K_S^0\pi^0$ (top) and $K_S^0 K_S^0$ (bottom) final states. Points with error bars are the data, the solid curve shows the result of the fit, long-dashed (red) is for the signal, dashed (blue) is for the total background and dotted (cyan) is for the peaking background.}
%\label{fig:total}
%\end{center}
%\end{figure}
%The $\Delta M $ distributions for $D^0 \to K_S^0 \pi^0$ and $D^0 \to K_S^0 K_S^0$ are shown in Fig.~\ref{fig:total}. 

We describe the $\Delta M $ distributions for $D^0 \to K_S^0 K_S^0$ and $D^0 \to K_S^0 \pi^0$ using the sum of two symmetric and one asymmetric Gaussian functions with a common most probable value. All the mode-dependent shape parameters are fixed from MC estimations, except for the mean and a common calibration factor for the symmetric Gaussians that accounts for a data-MC difference in the $\Delta M$ resolution. 

The backgrounds caused by processes with the same final state as the reconstructed modes, mainly $D^{0} \to K_S^0 \pi^+ \pi^-$ for the signal mode and $D^{0} \to \pi^{+}\pi^{-}\pi^{0}$ for the normalization mode, peak in the $\Delta M$ distribution. These peaking backgrounds are estimated directly from the data using the $K_S^0$ mass sidebands defined as $0.470$~GeV/$c^{2} < M_{\pi\pi} < 0.478$~GeV/$c^2$ and $0.516$~GeV/$c^{2} < M_{\pi\pi} < 0.526$~GeV/$c^2$. The peaking background has the same $\Delta M$ shape as the signal, and its yield is fixed based on the estimation described above to 267 events for $D \to K_S^0 \pi^+ \pi^-$ and 1923 events for $D^{0} \to \pi^{+}\pi^{-}\pi^{0}$. The combinatorial background shapes are modeled with an empirical threshold function $f(x)$ = $(x-m_{\pi})^{a}\exp[-b(x-m_{\pi})]$, where $m_{\pi}$ is the nominal charged pion mass, and $a$ and $b$ are shape parameters. 

\begin{figure}
\begin{center}
\includegraphics[width=9.0cm]{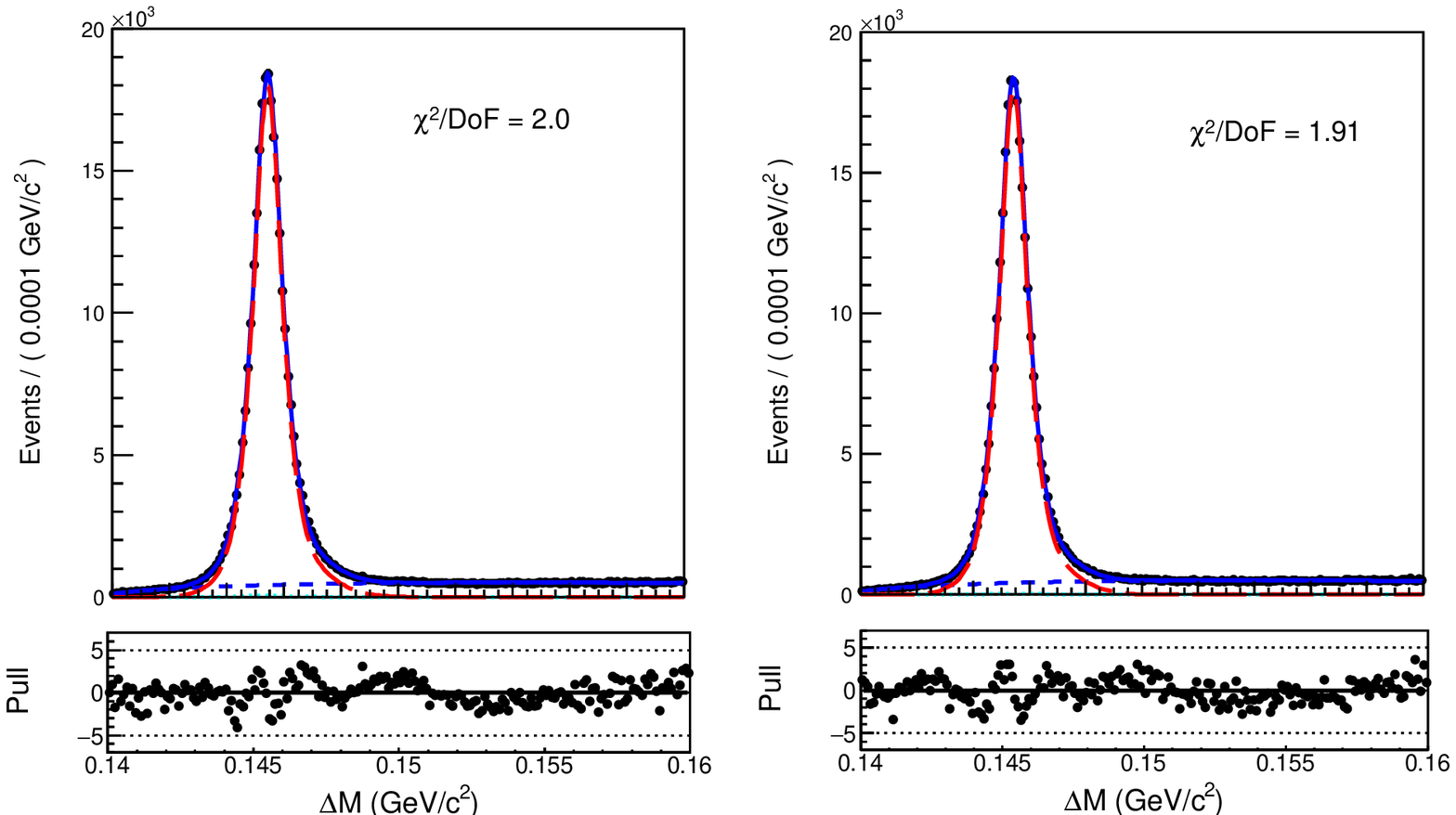}
\includegraphics[width=9.0cm]{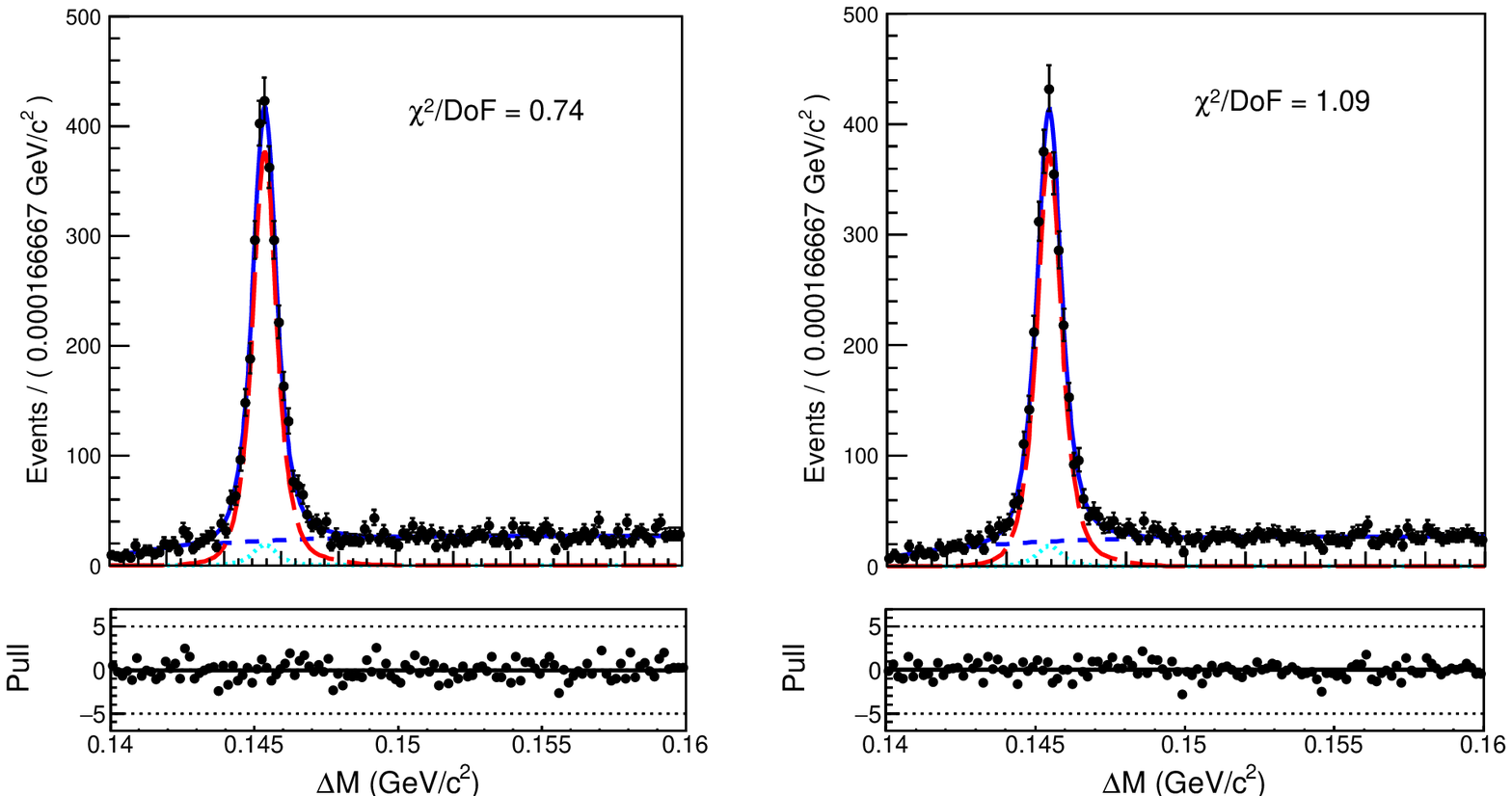}
\caption{Distributions of the mass difference $\Delta M$ for selected $D^{*+}$ (left) and $D^{*-}$ (right) candidates, reconstructed as $D^0 [K_S^0 \pi^0]\, \pi_{s}$(top) and $D^0 [K_S^0 K_S^0]\, \pi_{s}$ (bottom) decays. The points with error bars show the data, and the curves show the result of the fits with the following components: signal (long-dashed red), peaking background (dotted cyan), combinatorial background (dashed blue), and their sum (plain blue). The normalized residuals (pulls) and $\chi^2$/DOF, where DOF is the number of degrees of freedom, are also shown for each plot. } 
\label{fig:markern}
\end{center}
\end{figure}

An extended unbinned maximum likelihood fit to the two combined-charge $D^*$ $\Delta M$ distributions yields $5 399 \pm 87 $ $D^0 \to K_S^0 K_S^0$ events and $537 \ 360 \pm 833$ $D^0 \to K_S^0 \pi^0$ events. A simultaneous fit of the $\Delta M$ distributions for $D^{*+}$ and $D^{*-}$ (see Fig.~\ref{fig:markern}) is used to calculate the raw asymmetry in $D^{0}\rightarrow K_{S}^{0}K_{S}^{0}$. A similar procedure is followed for the $D^{0}\rightarrow K_{S}^{0}\pi^{0}$ sample. The signal and background shape parameters are common for both the particle and antiparticle. Both asymmetries in signal and background are allowed to vary in the fit. The value of $A_{\rm raw}$ for the peaking background in $D^{0}\rightarrow K_{S}^{0}\pi^{0}$ is fixed to zero, whereas its value in $D^{0}\rightarrow K_{S}^{0}K_{S}^{0}$ is fixed to the value obtained in the data for the $D^{0}\rightarrow K_{S}^{0}\pi^{0}$ signal. Here we assume that the peaking background in $D^{0}\rightarrow K_{S}^{0}\pi^{0}$ has zero net $A_{CP}$. The fitted values of $A_{\rm raw}$ for the $D^{0}\rightarrow K_{S}^{0}K_{S}^{0}$ and $D^{0}\rightarrow K_{S}^{0}\pi^{0}$ decay modes are $(+0.45 \pm 1.53)\%$ and $(+0.16 \pm 0.14)\%$, respectively. The resulting time-integrated $CP$-violating asymmetry in the $D^{0}\rightarrow K_{S}^{0}K_{S}^{0}$ decay is $A_{CP} = (-0.02 \pm 1.53) \%$.

For the branching fraction measurement, we use only the $D^{*+}$ candidates that have a momentum greater than 2.5~GeV/$c$ in the center-of-mass frame. This suppresses the component arising from $b\bar{b}$ events, and, hence, simplifies the efficiency estimation and controls the systematic uncertainty, which is the dominant uncertainty in this measurement. The $\Delta M$ fit yields $4 755 \pm 79$ $D^0 \to K_S^0 K_S^0$ decays and $475 \ 439 \pm 767$ $D^0 \to K_S^0 \pi^0$ decays. The selection efficiencies are (9.74 $\pm$ 0.02)\% and (11.11 $\pm$ 0.02)\%, respectively. Using Eq.~(\ref{eq:9}), we then obtain $\mathcal{B}(D^0 \rightarrow K_S^0K_S^0) / \mathcal{B}(D^0 \rightarrow K_S^0\pi^0)$ = (1.101 $\pm$ 0.023)\%. All quoted uncertainties are statistical. 

Table~\ref{tab:tablebf} lists various sources of systematic uncertainties in $A_{CP}$ and $\cal{B}$ of $D^{0}\rightarrow K_{S}^{0}K_{S}^{0}$. As the branching fraction measurement is a relative measurement, most of the systematic uncertainties common between the signal and normalization channel cancel. The uncertainties on the probability distribution function (PDF) parametrization are estimated by varying each fixed shape parameter by its uncertainty and repeating the fit. We independently vary the calibration factor for each Gaussian to account for different data-MC difference in the broad and narrow parts of the signal PDF. The systematic uncertainty is taken as the quadratic sum of the changes in the fitted results. 

The peaking background is estimated from the $K_S^0$ mass sidebands, and we fix the yield in the final fit using the scale factor between the signal region and sideband in the MC estimations after removing the signal contamination. We repeat the fit procedure by varying the fixed yield by its statistical error, and we take the difference between the resulting signal yield and the nominal value as the systematic uncertainty due to the fixed peaking background. We refit by varying the fixed $A_{\rm raw}$ by its statistical error and take the difference of the refitted and nominal results as the systematic uncertainty. The uncertainty due to fixing $A_{\rm raw}$ for the peaking component in both $D^{0}\rightarrow K_{S}^{0}K_{S}^{0}$ and $D^{0}\rightarrow K_{S}^{0}\pi^{0}$ is negligible. The dominant systematic uncertainty on $A_{CP}$ is from the uncertainty on the $A_{CP}$ measurement of the normalization channel, $D^{0} \rightarrow K_S^0 \pi^0$.
   
The systematic uncertainties on the reconstruction efficiency that do not cancel in the ratio to the normalization mode are those related to the reconstruction of the $K_S^0$ and the $\pi^0$. For both the MC estimations and data, the $K_S^0$ reconstruction efficiencies are estimated by calculating the ratio $R$  of the $D^0 \to K^0_S \pi^0$ signal yield extracted with and without the nominal $K^0_S$ requirements. Then, the double ratio $R_{\rm data}/R_{\rm MC}$= (98.57 $\pm$ 0.40)\% quantifies the possible difference between the data and simulations. We correct for the efficiency and assign a systematic uncertainty of 1.40\%. The tracking efficiency per track of $0.35\%$ is obtained from a large sample of $D^{*\pm} \rightarrow D^{0} \pi^{\pm}$, where the $D^{0}$ decays to $K_S^0\pi^{+}\pi^{-}$~\cite{pi0}. It is added linearly for the two daughters of the $K_S^0$ and combined with the above uncertainty, yielding $1.57\%$ for the systematic uncertainty due to $K_S^0$  reconstruction. There is a systematic uncertainty on the $\pi^{0}$ reconstruction efficiency. We obtain the corresponding data-MC correction factor (95.14 $\pm$ 2.16)\% from a sample of $\tau^{-} \rightarrow \pi^{-}\pi^{0}\nu_{\tau}$ decay~\cite{pi0}. We apply this correction and assign 2.16\% as a systematic uncertainty. Lastly, we take the uncertainty on the world-average branching fraction of the normalization mode $D^{0} \rightarrow K_S^0\pi^0$. These individual contributions are added in quadrature to obtain the total systematic uncertainty. 
 
\begin{table}[h]
\begin{center}  
\caption{Contributions to the systematic uncertainties of the measurements of the $CP$ asymmetry $A_{CP}$ (absolute errors) and branching fraction $\cal{B}$ (relative errors) for the $D^0 \to K_S^0 K_S^0$ mode.}

\begin{tabular}{l c c}  \hline\hline
Source &   $A_{CP}$ (\%) &$\cal{B}$ (\%)\\ \hline
$D^{0}\rightarrow K_{S}^{0}K_{S}^{0}$  PDF parametrization & $\pm 0.01$ &  $\pm 0.28$\\
$D^0 \to K^0_S \pi^0$ PDF parametrization& $\pm 0.00$ &  $\pm 0.23$  \\
$D^{0}\rightarrow K_{S}^{0}K_{S}^{0}$ peaking background & $\pm 0.01$ & $\pm 0.59$ \\
$D^0 \to K^0_S \pi^0$ peaking background & $\pm 0.00$ & $\pm 0.03$ \\
$K^{0}/\bar{K^{0}}$ material effects & $\pm 0.01$ & . . .\\
$K_{S}^{0}$ reconstruction efficiency      &  . . . &   $\pm 1.57$  \\
$\pi^{0}$ reconstruction efficiency     &  (. . .) &     $\pm 2.16$ \\
\hline
Quadratic sum of above & $\pm 0.02$ &   $\pm 2.76$ \\
\hline 
External input ($D^0 \to K_S^0 \pi^0$ mode) & $\pm 0.17$ & $\pm 3.33$ \\
\hline\hline
\end{tabular}
\label{tab:tablebf}
\end{center}
\end{table}

Using a data sample that corresponds to an integrated luminosity of 921 fb${}^{-1}$, we have measured the time-integrated $CP$-violating asymmetry in the $D^{0}\rightarrow K_{S}^{0}K_{S}^{0}$ decay to be

$$
 A_{CP} = (-0.02 \pm 1.53 \pm 0.02 \pm 0.17) \%,
$$

\noindent  where the first uncertainty is statistical, the second is systematic, and the third is due to the uncertainty on $A_{CP}$ of $D^0 \to K^0_S \pi^0$. From our measurement of the branching fraction ratio,

$$ 
 \frac{\mathcal{B}(D^0 \rightarrow K_S^0K_S^0)}{\mathcal{B}(D^0 \rightarrow K_S^0\pi^0)} = (1.101 \pm 0.023 \pm 0.030)\%,
$$ 

\noindent  we obtain the $D^{0}\rightarrow K_{S}^{0}K_{S}^{0}$ branching fraction as

$$
\mathcal{B}(D^{0}\rightarrow K_{S}^{0}K_{S}^{0}) = (1.321 \pm 0.023 \pm 0.036 \pm 0.044)\times 10^{-4},
$$ 
\noindent  where the first uncertainty is statistical, the second is systematic, and the third is due to the uncertainty on $\mathcal{B}$ of $D^0 \to K^0_S \pi^0$. 

 The $A_{CP}$ result is consistent with the SM expectation and improves the uncertainty with respect to the recent measurement of this quantity by LHCb~\cite{Aaij:2015fua} by about a factor of 4. Furthermore, the precision is already comparable to the theory prediction~\cite{Nierste:2015zra}. While the $\mathcal{B}$ result is consistent with the world average~\cite{pdg}, it is 2.3$\sigma$ away from a recent BESIII measurement~\cite{BESIII}. Both the $A_{CP}$ and $\mathcal{B}$ measurements are the most precise ones available for the $D^{0}\rightarrow K_{S}^{0}K_{S}^{0}$ mode. \\

%............................................................

We thank the KEKB group for the excellent operation of the accelerator; the KEK cryogenics group for the efficient operation of the solenoid; and the KEK computer group, the National Institute of Informatics, and the PNNL/EMSL computing group for valuable computing and SINET5 network support.  We acknowledge support from the Ministry of Education, Culture, Sports, Science, and Technology (MEXT) of Japan, the Japan Society for the Promotion of Science (JSPS), and the Tau-Lepton Physics Research Center of Nagoya University; the Australian Research Council; Austrian Science Fund under Grant No.~P 26794-N20; the National Natural Science Foundation of China under Contracts No.~10575109, No.~10775142, No.~10875115, No.~11175187, No.~11475187, No.~11521505 and No.~11575017; the Chinese Academy of Science Center for Excellence in Particle Physics; the Ministry of Education, Youth and Sports of the Czech Republic under Contract No.~LTT17020; the Carl Zeiss Foundation, the Deutsche Forschungsgemeinschaft, the
Excellence Cluster Universe, and the VolkswagenStiftung; the Department of Science and Technology of India; the Istituto Nazionale di Fisica Nucleare of Italy; the WCU program of the Ministry of Education, National Research Foundation (NRF) of Korea Grants No.~2011-0029457, No.~2012-0008143, No.~2014-R1A2A2A01005286,
No.~2014-R1A2A2A01002734, No.~2015-R1A2A2A01003280, No.~2015-H1A2A1033649, No.~2016-R1D1A1B01010135, No.~2016-K1A3A7A09005603, No.~2016-K1A3A7A09005604, No.~2016-R1D1A1B02012900, No.~2016-K1A3A7A09005606, No.~NRF-2013-K1A3A7A06056592; the Brain Korea 21-Plus program, Radiation Science Research Institute, Foreign Large-size Research Facility Application Supporting project and the Global Science Experimental Data Hub Center of the Korea Institute of Science and Technology Information;
the Polish Ministry of Science and Higher Education and the National Science Center; the Ministry of Education and Science of the Russian Federation and the Russian Foundation for Basic Research; the Slovenian Research Agency; Ikerbasque, Basque Foundation for Science and the Euskal Herriko Unibertsitatea (UPV/EHU) under program UFI 11/55 (Spain);
the Swiss National Science Foundation; the Ministry of Education and the Ministry of Science and Technology of Taiwan; and the U.S.\ Department of Energy and the National Science Foundation.

\end{document}